\documentclass[12pt,a4paper]{article}
\usepackage{amsthm,bm,amsmath,amssymb,epsfig,multicol,multirow}
\usepackage{natbib}
\usepackage{graphics,rotating,graphicx}
\usepackage{array,multirow,caption2}
\usepackage{amssymb}
\usepackage{amsthm}
\usepackage{bm,amsmath,multicol}
\usepackage{booktabs}

\renewcommand{\arraystretch}{1.5}

\def\mse{\hbox{MSE}}

\def\min{\hbox{min}}

\begin{document}

\noindent{\Large \textbf{Exploiting the Quantile Optimality Ratio  to
 Obtain\\ Better Confidence Intervals for Quantiles}}

\vspace{0.5cm}

\noindent LUKE A. PRENDERGAST

\noindent \textit{Department of Mathematics and Statistics, La Trobe
University, Melbourne, Australia, 3086}

\vspace{0.5cm}

\noindent ROBERT G. STAUDTE

\noindent \textit{Department of Mathematics and Statistics, La Trobe
University, Melbourne, Australia, 3086}
\vspace{1cm}

\noindent\textbf{ABSTRACT.  A standard approach to confidence intervals for quantiles requires good estimates of the quantile density. The optimal bandwidth for kernel estimation of the quantile density depends on an underlying location-scale family only through the quantile optimality ratio (QOR), which is the starting point for our results.  While the QOR is not distribution-free, it turns out that what is optimal for one family often works quite well for families having similar shape. This allows one to rely on  a single representative QOR if one has a rough idea of the distributional shape. Another option that we explore assumes the data can be modeled by the highly flexible generalized lambda distribution (GLD), already studied by others, and we show that using the QOR for the estimated GLD can lead to more than competitive intervals.  Confidence intervals for the difference between quantiles from independent populations are also considered, with an application to heart rate data.}

\vspace{0.5cm}

\noindent \textit{Key words:} coverage probability; density quantile; optimal bandwidth; quantile density

\clearpage
\newpage
\section{Introduction}

This work is motivated by the need for good interval estimates for quantiles when one has only a rough idea of the underlying shape, and  builds on substantial theory already in the literature.
Here we concentrate on asymptotic intervals of the simple form provided shortly in (\ref{ciquant}), usually adding and subtracting two standard errors to the quantile estimate. The challenge is
to obtain a good \lq distribution-free\rq\ estimator of the standard error, so as to obtain intervals for
quantiles with accurate coverage for all distributions having a similar shape.
There is also an extensive literature on completely distribution-free
confidence intervals for quantiles and, as a point of departure,
we refer the reader to \citet[Sec.2.6]{Serf-1980} for exact and asymptotic distribution-free intervals
based on order statistics and to \citet[Ch.29]{Das-2008} for bootstrap intervals.

Denote the {\em quantile function} associated with a cumulative distribution function $F$ by  $Q(u)\equiv F^{-1}(u)=\inf \{x:\ F(x)\ge u \}$, for $0< u< 1.$  Assuming $F $ has a positive derivative  $F'(x)=f(x)$ on its domain define $q(u)=Q'(u)=1/f(Q(u))$ to be the
  {\em quantile density function}  by \cite{par-1979}, and earlier dubbed the \lq sparsity index\rq \ by \cite{tukey-1965}.
 A point estimate of $Q(u)$ based on a sample $X_1,\dots ,X_n$ of size $n$ from $F$ is the sample quantile $\hat Q_n(u)=F_n^{-1}(u)$, where $F_n(x)$ is the usual empirical distribution function. Letting $ \tau ^2_u\equiv u(1-u)q^2(u) $, it can be shown \cite[Ch.7]{Das-2008} that the Studentized quantile is asymptotically normal:
$\sqrt {n}\{\hat Q_n(u) -Q(u)\}/\tau _u\to N(0,1)$ in distribution.
This leads to a large sample 100$(1-\alpha )$\% confidence interval for $Q(u)$ of the form
\begin{equation}\label{ciquant}
    \hat Q_n(u) \pm z_{1-\alpha /2}\,\frac{\tau _u}{\sqrt n\,}~,
\end{equation}
 where $z_\alpha =\Phi ^{-1}(\alpha )$ and
$\Phi $ is the standard normal distribution function.
 To make this confidence interval for $Q(u)$ distribution-free, one needs
to replace $\tau _u$ in (\ref{ciquant}) by a consistent estimator $\hat \tau _u$, which in turn requires a
consistent estimator of the quantile density $q(u)$. Thus there are two sources of error in the interval (\ref{ciquant}), that due to estimation of the center and that due to estimation of the width. \cite{shma-1990} have compared several  asymptotically optimal kernel estimators $\hat Q(u)$ of $Q(u)$ and found
\begin{small}
\begin{quote}
 ``... that apart from the extreme quantiles, there is little difference between various
 quantile estimators (including the sample quantile). Given the well-known distribution-free inference procedures (e.g., easily constructed confidence intervals) associated with the sample quantile, as well as the ease with which it can be calculated, it will often be a reasonable choice as a quantile estimator." - \cite{shma-1990}.
\end{quote}
\end{small}
Here we estimate the quantiles $x_u=Q(u)$ for $u\in [0.05, 0.95]$ using a linear combination of adjacent order statistics $\hat x_u$, the Type~8 version of the quantile command on the package \cite{R} recommended by \cite{HY&FA96}. Since estimation of the quantile density $q(u)$ is more
difficult, with $\mse [\hat q(u)]=O(n^{-4/5})$ and $\mse [\hat Q(u)]=O(n^{-1})$, the more important estimate in the confidence interval (\ref{ciquant}) is not the center, but the width of the interval, which requires an estimate of $q(u)$.

  \cite{jones-1992} makes a strong case for estimating $q(u)$ directly by kernel methods, rather than by taking the reciprocal of a kernel estimate of $f(x_u).$  It turns out that an  asymptotically optimal
choice of bandwidth for estimating $q(u)$ only depends on what we choose to call the {\em quantile optimality ratio} QOR$(u)=q(u)/q''(u)$, so this is the object of our attention in Section~\ref{sec:qor}. \cite{welsh-1988} suggests estimating $q(u)$ and $q''(u)$ separately and taking the ratio of these estimates to find the optimal bandwidth;
and \cite{cheng-2006} do so in computing the mean-squared error of estimators of $Q(u).$
While ideally one would consistently estimate the QOR at $u$, it turns out that one only
needs a rough estimate of it to obtain good confidence intervals for $x_u$.

Our goal is finding conceptually and computationally simple
 distribution-free confidence intervals for $x_u$, and
  propose using either one of the optimal bandwidth kernel estimators for a representative location-scale family, or to use an adaptive estimator for the generalized Tukey $\lambda $ families.  For the latter, our work is motivated by recent research from several authors. In particular  we briefly describe in Section~\ref{methods}
  the methods of \cite{SU09} along with those of \cite{soni-2012}.   We compare the finite sample performance of their confidence intervals with ours in Section~\ref{simulations} and also discuss extension to two-population comparisons.
 We conclude with an example in Section~\ref{simulations} and
 a discussion and summary in Section~\ref{summary}.

\section{Quantile optimality ratios}\label{sec:qor}

For background material on kernel density estimation see \cite{wand-1995}.

\subsection{Quantile density estimators based on optimal bandwidth}\label{sec:qordefn}

An appealing and simple to implement quantile density estimator is a kernel density
estimator which can be expressed as a linear combination of order statistics:
\begin{equation}
    \hat q(u)=\sum _{i=1}^nX_{(i)}\,\left \{k_b\left (u-\frac {(i-1)}{n}\right )
    -k_b\left (u-\frac {i}{n}\right )\right \},\label{qj}
\end{equation}
where $b$ is a bandwidth and $k_b(\cdot)=k(\cdot -b)/b$ for some kernel function $k$ which  is an even function
on $[-1,1]$ that has variance $\sigma ^2_k=\int x^2\,k(x)dx$ and roughness $\kappa =\int k^2(y)\,dy$. The asymptotic properties of this quantile density estimator
  have been studied by \cite{falk-1986}, \cite{welsh-1988} and \cite{jones-1992}, and
  the last of these gives the  asymptotic MSE of $\hat q(u)$ as:
\begin{equation}\label{asymMSE}
    \mse [\hat q(u)] =\frac {b^4 \sigma _k^4\{q''(u)\}^2}{4}+ \frac {\kappa \,q^2(u)}{bn}~.
\end{equation}
By differentiating (\ref{asymMSE}) with respect to $b$ one finds that  $\mse [\hat q(u)]$ has a minimum when the bandwidth $b(u)=A(u)/n^{1/5}$, where
\begin{equation}\label{bwconstant}
    A(u) = \left (\frac {\kappa }{\sigma _k^4}\right )^{1/5}\; \left \{ \frac {q(u)}{q''(u)}\right \}^{2/5}~.
\end{equation}
Therefore an asymptotically optimal choice of bandwidth (\ref{bwconstant}) for estimating $q(u)$ only depends on the underlying distribution through the QOR$(u)=q(u)/q''(u)$.
Note that if $F_{a,b}(x)=F((x-a)/b)$, for unknown $a$ and $b>0$, is the location-scale family generated by $F=F_{0,1}$, then the quantile function
for $F_{a,b}$ is $Q_{a,b}(u)=a+b\,Q(u)$ and the quantile density is $q_{a,b}(u)=b\,q(u)$; thus the quantile density is location-invariant and scale equivariant, and the QOR is both location and scale invariant.

\subsubsection*{Remarks on derivation of the QOR}The first two derivatives of the quantile density $q$ are:
\begin{eqnarray}\label{eqn:quantilederivs} \nonumber
  q'(u) &=& -\frac {f'(x_u)}{f^3(x_u)}\ =\ -f'(x_u)\,q^3(u)\\
  q''(u) &=& 3\{f'(x_u)\}^2\,q^5(u)-f''(x_u)\,q^4(u) ~.
\end{eqnarray}
In many cases, $f'(x)=g(x)f(x)$. (For example, in the Pearson systems of distributions, see \citet[Ch1.]{J-K-B-1994}, $g(x)$ is a rational function whose numerator is linear in $x$ and denominator is quadratic in $x$.)  For such $f$, we have $f''(x)=\{g'(x)+g^2(x)\}\,f(x)$ so that from (\ref{eqn:quantilederivs}) one finds $q'(u)=-g(x_u)q^2(u)$ and $q''(u)=\{2g^2(x_u)-g'(x_u)\}q^3(u); $ thus
the quantile optimality ratio becomes:
\begin{equation}\label{eqn:quantileratio}
  \text{QOR}(u)=\frac {q(u)}{q''(u)} = \frac {f^2(x_u)}{\{2g^2(x_u)-g'(x_u)\}\,} ~.
\end{equation}
In this case $\sqrt {\text{QOR}(u)}$ is the product of the {\em density quantile} function $f(x_u)=1/q(u)$ defined by
\cite{par-1979} and a function depending only on $g$ and its derivative composed with the quantile function.
Perhaps it is worth noting that the score function $J(u)=-f'(x_u)/f(x_u)=-g(x_u)$ of classical nonparametric statistics is defined in terms of $g(x_u)$.

It is informative to examine some QOR's before looking at specific methods for estimation of $x_u$ in Section~\ref{methods}, because they are much better behaved then we expected.

\subsection{Examples of the quantile optimality ratio}\label{sect:qorexs}

The quantile density for the uniform distribution is a positive constant, so its  derivatives are 0, and the QOR$(u)=+\infty $ for all $0<u<1$; thus the optimal bandwidth result does not apply in the uniform case.

\subsubsection*{Normal.} 
\begin{figure}[t!]
\centering
\includegraphics[scale = 0.7]{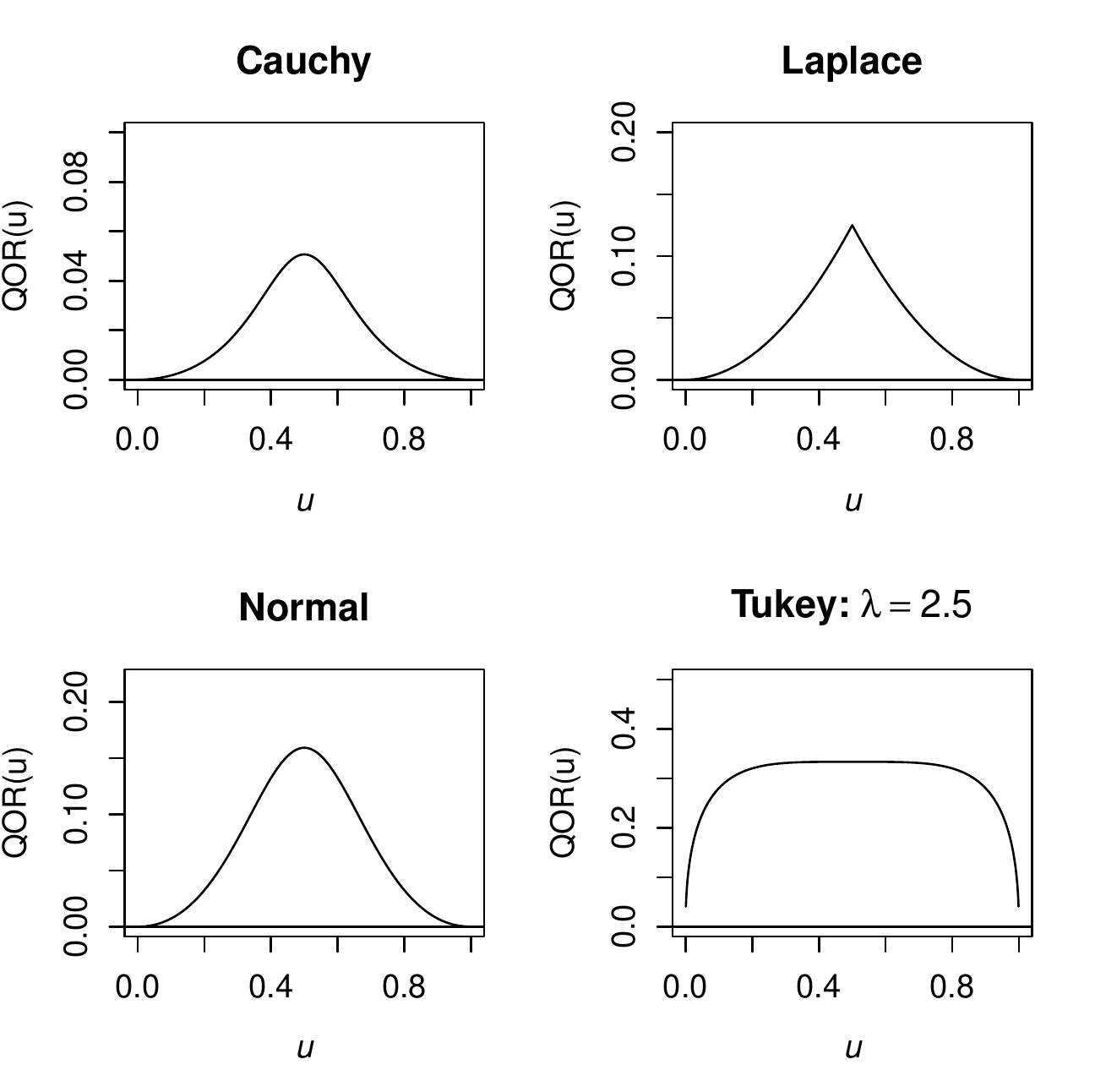}
\caption{Plots of the QOR  functions for some symmetric distributions. Formulae for the QORs can be found in Table \ref{table:QOR}.}\label{fig1}
\end{figure}

Letting $z_u=Q_\Phi (u)=\Phi ^{-1}(u)$, the quantile density function is $q_\Phi (u)=1/\varphi (z_u),$ with first two derivatives $q_\Phi '(u)=z_u/\varphi ^2(z_u)=Q_\Phi (u)\,q_\Phi ^2(u)$ and $q_\Phi ''(u)=q_\Phi ^3(u)(1+2Q_\Phi ^2 (u)).$  The graph of the QOR is in the bottom left
plot of  Figure~\ref{fig1}. This ratio can be written $\sqrt 2\,\varphi (\sqrt 2\,x_u )/\{\sqrt \pi\,(1+2x_u^2)\}.$   As an aside, it also has a simple approximation $QOR_\Phi (u)\approx 0.4\;\,\varphi(6(u-0.5))$ for $u$ real, which perhaps explains why it appears to
be a normal density.

In Figure~\ref{fig1} we also plot the QOR for Cauchy, Laplace and Tukey$(\lambda)$ distribution with $\lambda = 2.5$.  For symmetric distributions that are unimodal and positive for all $x$ the graphs of $q(u)$ and $q''(u)$ are U-shaped, symmetric about $1/2$ and unbounded, while  $q'(u)$ is skew-symmetric about 1/2 and unbounded. However, the ratio QOR$=q(u)/q''(u)$ is bounded and even recognizable:   a first glance at Figure~\ref{fig1} suggests that we are plotting  density quantiles of some well-known distributions, when in actual fact they are the plots of the graphs of the quantile optimality ratios.

\begin{figure}[h!]
\centering
\includegraphics[scale = 0.7]{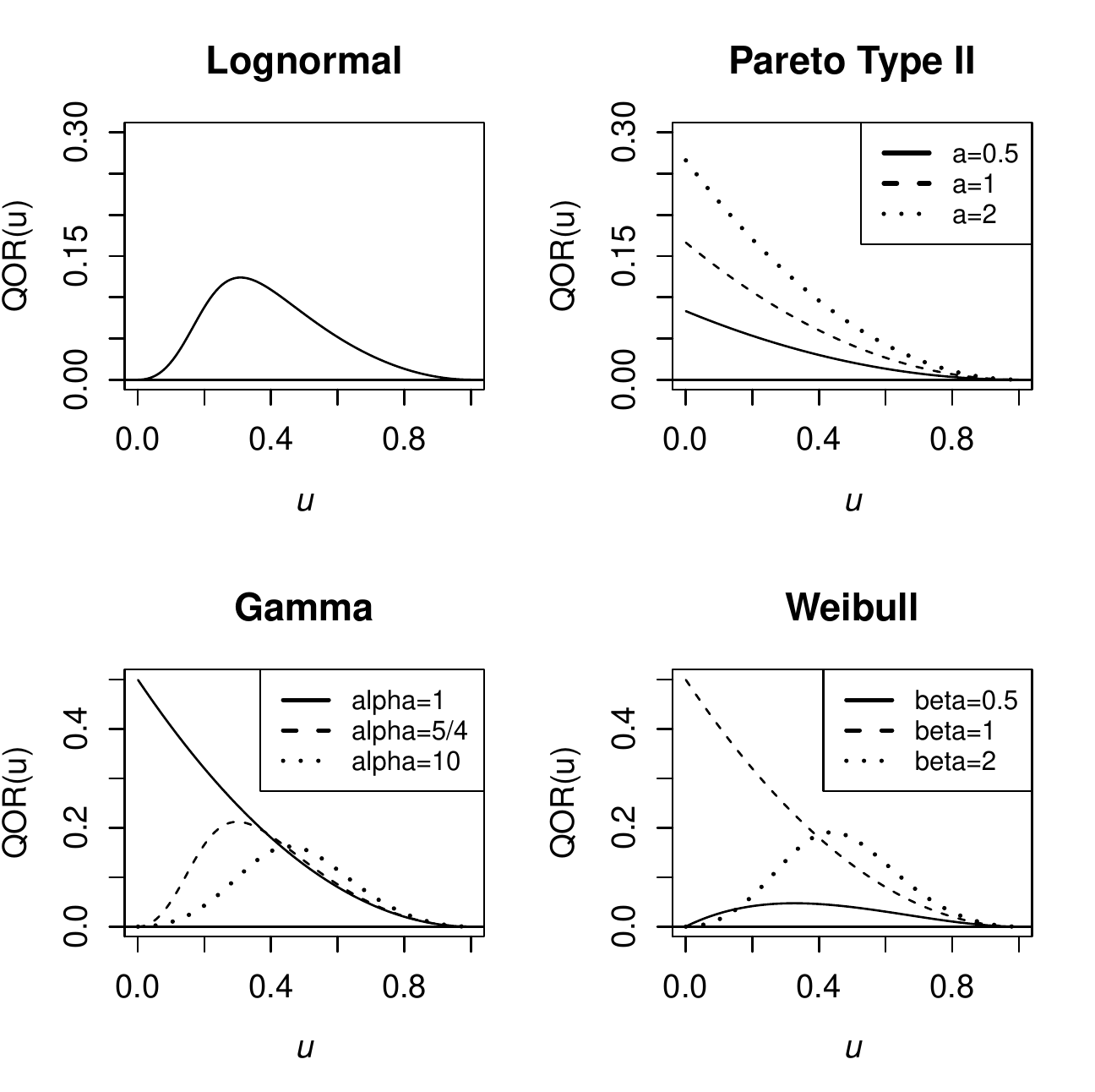}
\caption{Plots of the QOR  functions for some asymmetric distributions.}\label{fig2}
\end{figure}

\subsubsection*{Lognormal.}
For $x>0$ let $F(x)=\Phi (\ln(x)) $ be the log-normal distribution function, and recall $q_\Phi (u)=1/\varphi (z_u)$ is the quantile density of the normal distribution.
It follows that $f(x)=F'(x)=\varphi(\ln(x))/x.$ Therefore $x_u=Q(u)=\exp \{z_u\}$ for $0<u<1$ and the quantile density and its first two derivatives are
\begin{eqnarray} \label{lnormqden1}
  q(u) &=&  \frac {x_u}{\varphi (z_u)}\;=\; Q(u)\,q_\Phi (u)\\ \nonumber
  q'(u) &=& q(u)\,q_\Phi (u)+ Q(u)\,q_\Phi '(u)\\ \label{lnormqden3}
  q''(u)&=& q'(u)\,q_\Phi (u)+  2q(u)\,q_\Phi ' (u)+Q(u)\,q_\Phi ''(u)
\end{eqnarray}
The graph of the quantile optimality ratio  QQR$(u)$ is shown in the top left plot of Figure~\ref{fig2}.  For comparison, also appearing in Figure \ref{fig2} are the QORs for the Pareto type II, the Gamma and the Weibull distributions, each with three parameter choices.

\subsubsection*{Generalized Lambda Distribution.}

We adopt the parameterisation of \cite{fmkl-1988} (often referred to as the FKML parameterisation), which has
quantile function determined by  a location parameter $\lambda _1$, an inverse scale parameter $\lambda _2>0$ and two shape parameters $\lambda _3$ and $\lambda _4:$
\begin{eqnarray} \label{eqn:Qgld}
Q(u) = \lambda_1 +\frac{1}{\lambda _2}\,\left \{\frac {u^{\lambda _3} -1}{\lambda _3} - \frac {(1-u)^{\lambda_4 }-1}{\lambda_4 }\right \}~.
\end{eqnarray}
The quantile optimality ratio QOR$(u)$ is location and scale free so without loss of generality we can
 take $\lambda _1=0$ and $\lambda _2=1$ in (\ref{eqn:Qgld}). The QOR is easily found and shown in Table~\ref{table:QOR} (see GLD-FKML); details are left to  the reader.  The RS parameterisation \citep[RS][]{RS} is sometimes also used and this is also included in the table.

This generalized Tukey family of distributions can be used to approximate a large number of others, as described in \cite{fmkl-1988},   \cite{kardud-2000} and \cite{gilch-2000}.
Thus an appealing approach is to estimate the parameters $\lambda _3$ and $\lambda _4$ and then use
the QOR for this family to estimate the optimal bandwidth when estimating $q(u)$ and finding a confidence interval for $x_u=Q(u).$

\renewcommand{\arraystretch}{1.8}
\begin{table}[h!t]
\begin{small}
\begin{tabular}{ll}
  \hline
  Family &  QOR$(u) $ \\ \hline
  Cauchy  &  $\displaystyle \left [(2\pi \sqrt {3})\{1+3x_u^2\}\right ]^{-1}$\\
  Laplace &  $\displaystyle  \begin{cases}{}
              u^2/2,\qquad \quad u< 0.5 \\
             (1-u)^2/2,\; \; u\geq 0.5 .\\
            \end{cases} $\\
 Normal & $\displaystyle  \frac{\sqrt 2\,\varphi (\sqrt 2\,z_u )}{\sqrt \pi\,(1+2z_u^2)}$ where $z_u\equiv \Phi ^{-1}(u)$.\\
 Lognormal &  Use formulae (\ref{lnormqden1}) and (\ref{lnormqden3}). \\
 Logistic &  $\displaystyle  \frac {\{ u(1-u)\}^2}{2\{(1-u)^3+u^3\}}$ \\
Bimodal &     0.25 \\
 Tukey$(\lambda)$ &   $\displaystyle
  \frac{u^{\lambda -1}+(1-u)^{\lambda -1}}{(\lambda -1)(\lambda -2)\{u^{\lambda -3}+(1-u)^{\lambda -3}\}}$\\
   Pareto$(a)$   &   $\displaystyle \frac{a^2(1-u)^2}{(1+a)(1+2a)}$\\
  Gamma $(\alpha )$   &   See (\ref{gammaqor}) in the Appendix. \\
   Exponential   &   $\displaystyle    (1-u)^2/2$\\
    Weibull $(\beta )$   & See (\ref{weibullqor}) in the Appendix. \\

    GLD-FKML$(\lambda _3,\lambda _4)$&  $\displaystyle \frac{u^{\lambda_3-1}+(1-u)^{\lambda_4-1}}{u^{\lambda_3-3}(\lambda_3-2)(\lambda_3-1) + (1-u)^{\lambda_4-3}(\lambda_4-2)(\lambda_4-1)}$\\
  GLD-RS$(\lambda _3,\lambda _4)$ & $\displaystyle \frac{\lambda_3u^{\lambda_3-1}+\lambda_4(1-u)^{\lambda_4-1}}{u^{\lambda_3-3}(\lambda_3-2)(\lambda_3-1)\lambda_3 + (1-u)^{\lambda_4-3}(\lambda_4-2)(\lambda_4-1)\lambda_4}$\\
  \hline
\end{tabular}
\end{small}
\caption{\footnotesize \em Quantile optimality ratios (QORs) for various distributions.  The normal, lognormal and GLD QORs are discussed in Section \ref{sect:qorexs} with further examples in the Appendix.}\label{table:QOR}
\end{table}
\renewcommand{\arraystretch}{1.5}

\subsubsection*{Other examples.}

 More examples of the QOR  are provided in Table~\ref{table:QOR} and  derivations of these QORs  are  in the Appendix (Section \ref{appendix}).

\section{Methods for interval estimates of quantiles}\label{methods}

 In the previous section we showed that optimal bandwidths only depend on the QOR, and that the latter can vary greatly in shape, depending on the underlying location-scale family.  Ideally one would find a good distribution-free estimate of the QOR, and use that to choose an approximately optimal bandwidth that gives good finite sample results. However, attaining this ideal is not necessary.

  After inspection of the data through histograms or density plots, one is often willing to assume something about the underlying distribution, such as unimodality and symmetry on infinite support or unimodality and support on a half-open interval $[0,\infty ).$  In the former case, choosing the optimal Cauchy, say, bandwidth can lead to relatively small MSEs of finite sample estimates of $q(u)$ for
data generated from Cauchy or other symmetric unimodal distributions. In the latter case, on can assume a Pareto model, estimate the shape parameter, and then proceed to estimate the QOR and optimal QOR bandwidth for this model to obtain the estimate of $q(u)$. The optimal QOR for the log-normal distribution can lead to good coverage of intervals for quantiles for many unimodal skewed distributions on $[0,\infty )$.

 Another possibility examined here is restricting attention to a larger parametric family of distributions, such as the four-parameter generalized lambda model, estimate its parameters and then proceed to estimate $q(u)$ as though this were the true distribution.

\subsection{Methods based on optimal QOR bandwidths}

These methods follow the \cite{par-1979}, \cite{falk-1986}, \cite{welsh-1988} and \cite{jones-1992} method of estimating $q(u)$ by $\hat q(u)$ as defined in (\ref{qj}) with an optimal bandwidth  $b(u)=A(u)/n^{1/5}$, where $A(u)$ is determined by (\ref{bwconstant}) and the Epanechnikov kernel. Thus $A(u)=1.718\; \text{QOR}(u)^{2/5}$ where the optimality ratio QOR$(u)$ depends on a fixed location-scale family, see Table~\ref{table:QOR}.

\subsubsection*{Method A:\quad Representative models.}

For symmetric unimodal densities with infinite support, a single QOR determined by the Cauchy distribution, say, can lead to  good interval estimators of $x_u=Q(u)$ of the form (\ref{ciquant})
for many of them, and lead to reliable confidence
intervals, as will be shown in Section~\ref{study1}.

Similarly, if a density is known to have a shape that is skewed to the right on $[0,\infty)$,  the QOR that is log-normal for this shape can lead to good interval estimators of quantiles for similarly shaped  distributions, even Pareto and exponential models, see Section~\ref{study2}.

\subsubsection*{Method B: \quad Simple adaptive parametric models.}

In the case of symmetric unimodal densities one could assume the 3-parameter symmetric Tukey$(\mu , \sigma, \lambda )$ family, and
estimate $\lambda $ with the data before choosing the QOR function that is optimal for this model to obtain the bandwidth, estimate $q_{\hat \lambda }(u)$ and then find the associated interval for  $x_u.$  See Ch. 7 of \cite{gilch-2000} for a discussion and references.

In the case of skewed-right densities on $[0,\infty)$,
if one is willing to assume for simplicity  a Pareto type II model with known scale 1, then an adaptive procedure which first estimates the shape parameter $a$ by the maximum likelihood estimate and
then uses the optimal bandwidth for the estimated model is an option.
The QOR$_a(u)=a^2(1-u)^2/\{(1+a)(1+2a)\}$ and the MLE $\hat a= n/\sum _i\ln (1+X_i). $ The performance of this estimator is studied in Section~\ref{study2}. Extensions to the more
practical case of unknown scale and
shape parameters is possible, see Ch.20 of \cite{J-K-B-1994} and the R package {\tt maxlik}.

\subsection{Methods based on the generalized lambda distribution}
It is tempting to restrict attention to the 4-parameter generalized lambda distribution (GLD) because it approximates a wide range of symmetric and asymmetric  distributions; this model requires implicitly defined estimates of
the parameters, now possible with most software packages. This model has two drawbacks: an explicit expression for the estimated density is not available except in special cases, and if the underlying distribution is not a member of the GLD family the methods will not be consistent for estimating the quantiles.  Methods
C,D and E will be compared in Section~\ref{study3}.

\subsubsection*{Method C}

\cite{SU09} suggests taking $f$ to be the density of the Generalized Lambda Distribution (GLD) which depends on the location parameter $\lambda_1$, inverse scale parameter $\lambda_2$ and the shape parameters $\lambda_3$ and $\lambda_4$.    Given estimates of $\lambda_1,\ldots,\lambda_4$ which give rise to estimated GLD quantile $\hat Q_C(u)$, estimated GLD density $\widehat{f}_C$ and estimated standard error
$\hat \tau _C/\sqrt n= \sqrt{u(1-u)}\;\widehat{f}_C(\widehat{Q}_C(u))/\sqrt n\,$ of the GLD quantile density, one
can construct a confidence interval (\ref{ciquant}) for $x_u$.  This approach is called  the Normal-GLD method
 by \cite{SU09} and we will call it Method C.

\subsubsection*{Method D}

\cite{SU09} also introduced a method for estimating $x_u=Q(u)$ based on the estimated GLD cumulative distribution
function $\widehat{F}$. Let $m=\text{floor}(nu)$  and
 $F_{\cal B}(\cdot;m+1,n-m)$ be the cumulative distribution function for the Beta ${\cal B}(m+1,n-m)$ distribution.
One then defines a $1-\alpha $ confidence interval $[L,U]$ by taking the solutions to $\alpha /2= F_{\cal B}(\widehat{F}(L); m+1,n-m)$ and $1-\alpha /2= F_{\cal B}(\widehat{F}(U); m+1,n-m).$ Such root-finding is routine on most statistical
packages. For justification of this approach, see \cite{SU09}, who calls this the
Analytical-GLD approach; here we call it Method D.
\cite{SU09} provided simulated evidence of typically improved performance of Methods C and D when compared to popular bootstrap counterparts.

\subsubsection*{Method E}

 In the spirit of Method~B we propose an adaptive GLD method
that first estimates the 4 parameters of the GLD distribution and employs the optimal QOR bandwidth for this estimated
model in finding the confidence interval. A caveat is that  any adaptive approach to bandwidth selection such as Method E  may not minimize MSE as well as  expected because the bandwidth is now random and the derivation of the optimal bandwidth assumes it only depends of $n$, $u$, the kernel and the underlying distribution.  It could turn out that using an optimal bandwidth for a fixed distribution with shape near that of the underlying distribution leads to better finite sample results than one which estimates unknown parameters first.

\subsection{Nonparametric methods}\label{soni}

Methods F,G and H below  are due to \cite{soni-2012} and are examined briefly in Section~\ref{study4}.

\subsubsection*{Method F}

The \lq reciprocal\rq\  approach to estimating $q(u)=1/f(Q(u))$ is to first estimate $Q(u)$ by \begin{equation}\label{Qhat}
    \hat Q(u)=\sum _iX_{(i)}\int _{(i-1)/n}^{i/n}k_b(u-y)\,dy
\end{equation}
and $f(x)$ by $\hat f(x)=(\sum _ik_b(X_i-x))/n$, where $b=b_n(u)$ is the bandwidth.
\cite{jones-1992} denotes this estimator $\tilde q(u)=1/\hat f(\hat Q(u))$ and
\cite{SO12} names it $q_n^{j1}(u).$ Here we call it $\hat q_F(u)$. Method~F for finding a confidence
interval for $x_u=Q(u)$ consists of substituting $\hat q_F(u)$ for $q(u)$ in (\ref{ciquant}).

\subsubsection*{Method G}

Rather than estimate $f(Q(u))$ and taking its reciprocal, \cite{jones-1992} proposed directly estimating $q(u)$
by (\ref{qj}) which he called $\hat q(u)$. \cite{SO12} denoted it $q_n^{j2}(u)$ and hereafter we call it
$\hat q_F(u).$   Method G  consists of substituting $\hat q_G(u)$ for $q(u)$ in (\ref{ciquant}).
The main conclusion of the asymptotic MSE results of  \cite{jones-1992} is that Method G is preferable to Method F, except perhaps for quantiles in the center. \cite{SO12} assert on the basis of their simulations that the reverse is true.

\subsubsection*{Method H}
\cite{SO12} proposed Method~H, which consists of substituting $\hat q_H(u)$ for $q(u)$ in (\ref{ciquant}), where
\begin{equation}
\widehat{q}_H(u)=\frac{1}{n}\sum^n_{i=1}\frac{k_b\left(S_i-u\right)}{\widehat{f}(X_{(i)})}\label{qs}
\end{equation}
 and $S_i$ is the proportion of observations less than or equal to $X_{(i)}$, $k_b(\cdot)=k(\cdot /b)/b$ is a kernel function and $\hat f$ is the usual kernel density estimator of $f$ defined in Method A.

\cite{SO12} show via simulations that for certain distributions and
selected choices of $u$ ranging from 0.2 to 0.9 that $\hat q_H$ was superior to $\hat q_F$ and $\hat q_G.$
Interestingly, they did not try to optimize the bandwidth $b=b_n(u)$ but chose it  arbitrarily to be a constant $0.15,$ $0.19$ and $0.25$ which led to similar results, so they only reported those for $0.19.$

\begin{figure}[th!]
\begin{footnotesize}
\centering
\includegraphics[scale = 0.7]{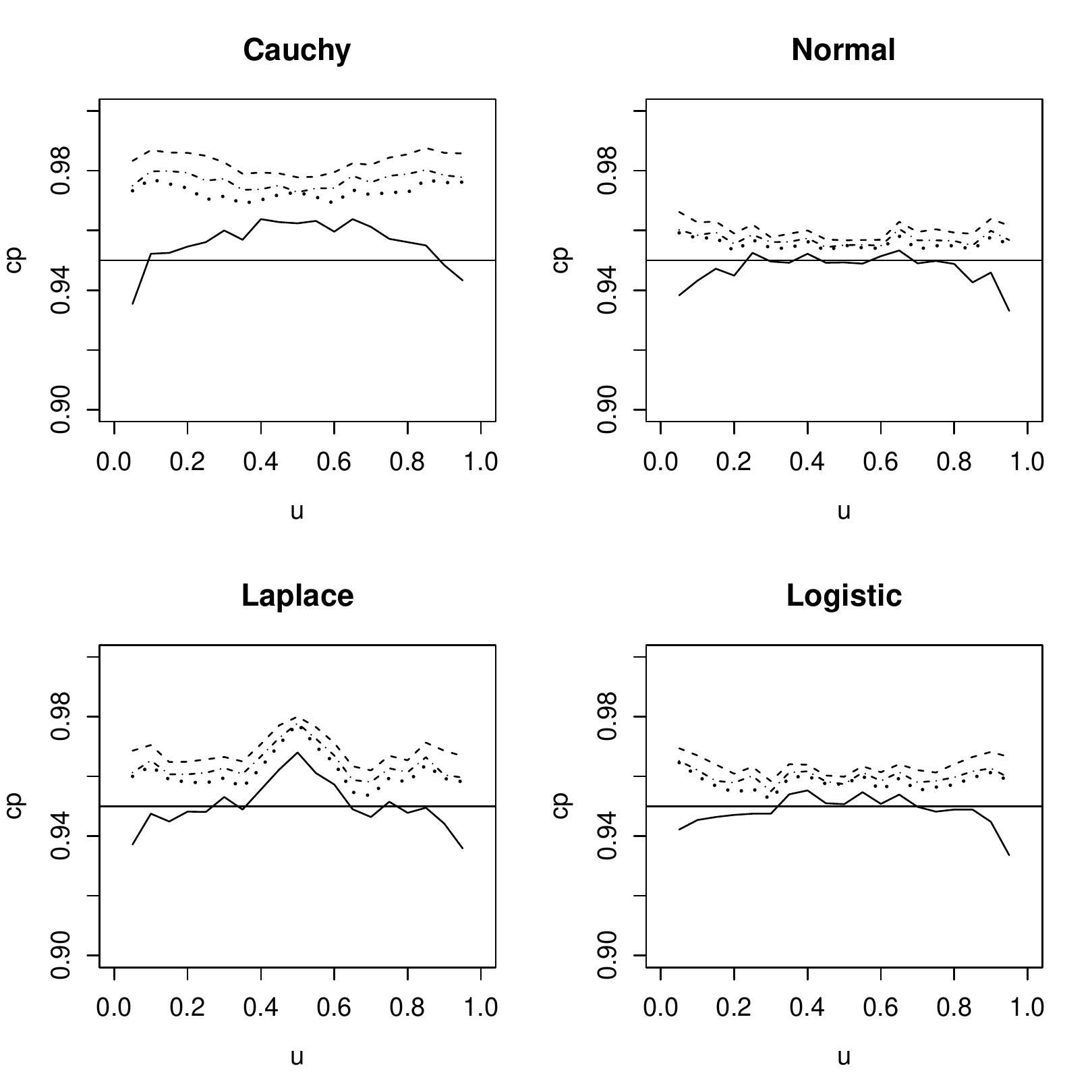}
\caption{{\bf n=400} \quad \em Estimated coverage probabilities for $x_u=Q(u)$ for $u=0.05:0.95/0.05$ based on 10,000 replicates  of four optimal 95\% confidence intervals. The intervals are based on optimal QOR's for the Cauchy (solid lines), the normal (dashed lines), the Laplace (dotted lines) and the logistic (dot-dashed) lines. The four plots are for data generated assuming these
respective models.\label{figure3}}
\end{footnotesize}
\end{figure}

\section{Simulation Studies}\label{simulations}

All studies that follow were carried out with the R statistical software package \citep{R}.  Specifically, to obtain the numerical maximum likelihood estimates for the GLD distribution, we used the \textit{gld} package \citep{KI&DE14}.  The package also includes many other estimators and the reader is referred to \cite{CO&ME15} for a discussion of the performance of many available estimators.  Additionally, other packages are also available that may be used to estimate the GLD parameters \citep[for e.g., the \texttt{GLDEX} package][]{SU07R}.

In this section we present several simulation studies that emphasize the usefulness of the optimal QOR approach in many situations and provide additional evidence that Method C of \cite{SU09} can provide very good results, in particular when $n$ is small. While relatively small bias and standard errors of $\hat q_n(u)$ are desirable, our primary
goal is good coverage probability  of the associated confidence intervals $[L,U]$ defined in (\ref{ciquant}) and secondary
goal of small  standardized widths $n^{1/2}(U-L);$ thus these quantities will be estimated by simulations and recorded.

\subsection{Study 1:  Performance of optimal bandwidth estimators  for symmetric unimodal $F$ with infinite support}\label{study1}
Here we compare four optimal estimators for the normal, logistic, Laplace and Cauchy quantile functions $q(u)$, and the
coverage and widths of the associated large sample 95\% confidence intervals (\ref{ciquant}) for $x_u=Q(u)$, for $u\in [0.05,0.95].$  All four intervals will be compared for data generated from all four distributions, to see if any of the
intervals gives acceptable results for all of them.  Preliminary examination of biases of these optimal $\hat q(u)$s reveals that all are biased upwards over this range of study  $u\in [0.05,0.95],$ with the optimal for Cauchy  being the least biased.  The standard errors of the  $\hat q(u)$s are very similar, and so are the average widths of the associated intervals.  For $n=100$, the average width of any of these intervals is approximately one
for $u=0.5$, and grows to 4 at $u=0.15$ (or $u=0.85$) and is 10 for $u=0.05$ (or $u=0.95$). In any case, for each $u$
the intervals decrease in size at the rate of $n^{-1/2},$ so these facts should be kept in mind in deciding whether a given
sample size is adequate for estimating a given quantile $x_u$ to a given accuracy.

In Figure~\ref{figure3} a comparison of the empirical coverages of these intervals is shown for sample size $n=400$ from each of the Cauchy, normal, Laplace and logistic distributions.  The confidence
interval based on the optimal Cauchy estimator for the quantile density function performs best, with coverage probability near the nominal value for all $u\in [0.1,0.9]$ for all four generated data sources.  The other three intervals have
conservative coverage over this range; they possibly could be improved by making finite sample adjustments.
   If $n$ is reduced to 100, simulations (not shown) demonstrate that only the optimal for Cauchy intervals have coverage near 95\% for these 4 settings, and then only over the range $[0.2,0.8]$; the other intervals are far too conservative to be of practical value.   For $n=200$, the range of useful coverage of the optimal Cauchy intervals is
 $[0.15,0.85]$; and for $n=800$ this range extends to  $[0.05,0.95]$.

In addition, the same optimal QOR for Cauchy interval gives close to 95\% coverage for $u\in [0.1,0.9]$ when sample sizes are $n=400$ from the symmetric Tukey distribution and $\lambda $ ranging from $-1$ to 5, although confidence coefficients
fall slightly below the nominal 95\% for the uniform cases $\lambda =1$ and $\lambda =2$. A thorough analysis, including
comparison with an optimal for the Tukey $\lambda $ family, with $\lambda $ estimated from the data, would be of interest but is beyond the scope of this work.

   In summary, the optimal for Cauchy QOR method is recommended for symmetric unimodal
   distributions with infinite support, and this is Method~A under such an assumption.

\subsection{Study 2:  Performance of optimal bandwidth estimators for skewed $F$ with support  on $[0,+\infty)$}\label{study2}

In the second study we generated data from skewed models often postulated for positive income data:\quad the lognormal, exponential, and Type II Pareto models with shape parameters $a=1$ and $a=2.$ For these four
models we compare the  confidence intervals associated with the optimal QOR estimators of $q(u)$ for the first
three of these models and  with those
of an adaptive method that first estimates the Pareto shape parameter $a$ by the MLE
$\hat a= n/\sum _i\ln (1+x_i)$ and then uses the optimal bandwidth QOR$_{\hat a}(u)$.

\vspace{0.2cm}

\noindent {\bf Important boundary correction:}
{\em Preliminary simulations showed that none of the above intervals had satisfactory coverage
for smallish $u<0.2$ because the ``optimal" bandwidths were extending beyond the
lower boundary of $[0,+\infty )$.  This problem was successfully resolved by taking the
bandwidth to be $b=\min \{u,b\}$. Hereafter we make this boundary correction to optimal
QOR bandwidths for distributions with support $[0,+\infty )$.}

\vspace{0.2cm}

\begin{figure}[th!]
\begin{footnotesize}
\centering
\includegraphics[scale = 0.7]{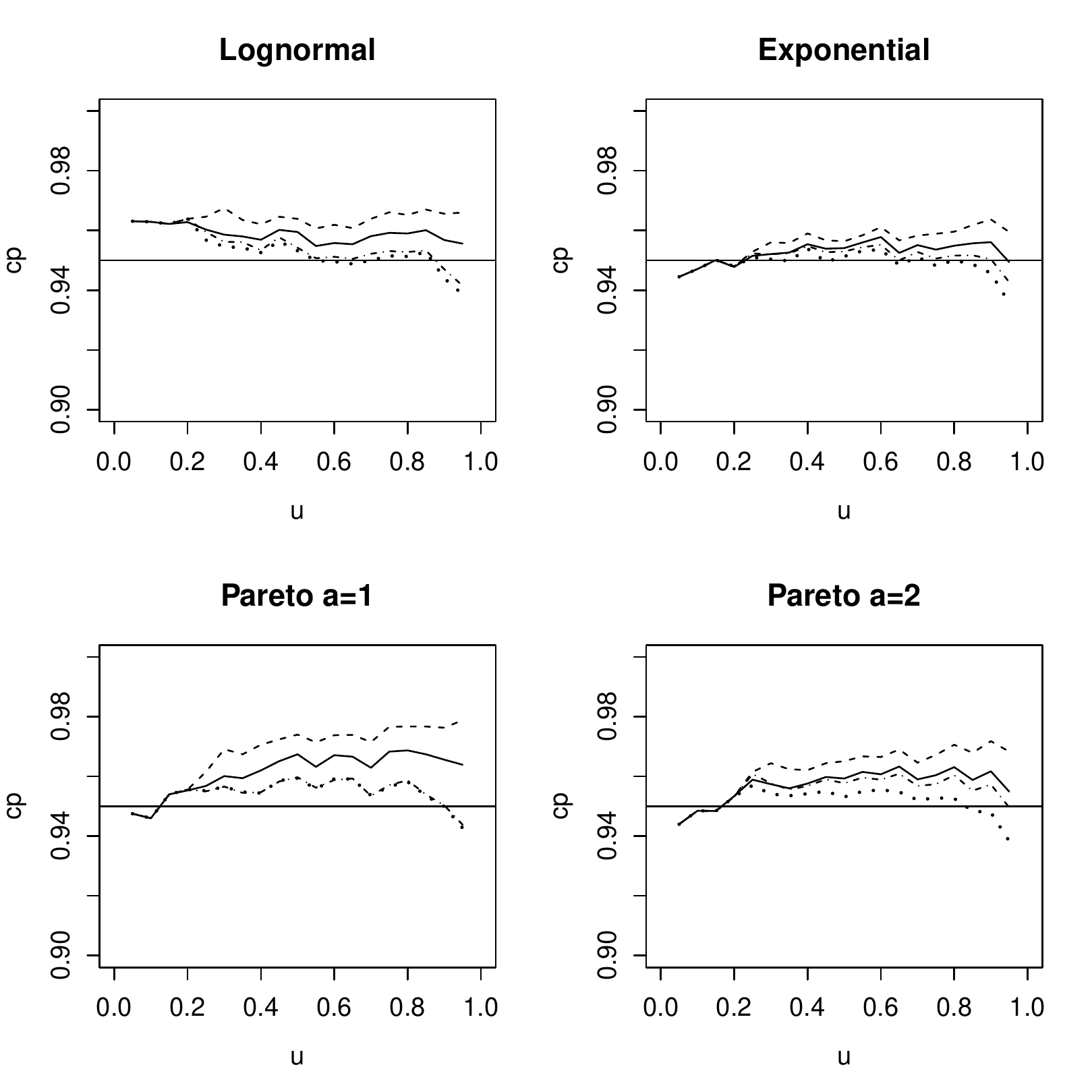}
\caption{{\bf n=400} \quad \em Estimated coverage probabilities based on 10,000 replicates of four  95\% confidence intervals for $x_u=Q(u)$, based respectively on optimal QOR's for the Lognormal (solid lines), the exponential (dashed lines), the Pareto$(a=1)$ (dotted lines) and the adaptive Pareto, which uses the MLE $\hat a$ to determine the QOR, (dot-dashed) lines. The four plots are for data generated from the Lognormal, exponential, Pareto$(a=1)$ and Pareto$(a=2)$.\label{figure4}}
\end{footnotesize}
\end{figure}

One can see from Figure~\ref{figure4} that when
$n=400$ no interval is uniformly better in terms of coverage than the others, and all have
similar coverage for small $u$ because of the boundary correction.
In general, the optimal QOR for exponential intervals have too conservative coverage, while the other three intervals perform satisfactorily for $u\in [0.05,0.95].$

In practice an adaptive Pareto model will also require estimation of scale, so we also
tried using maximum likelihood estimates for unknown scale and shape, but found this
approach unsatisfactory because the likelihood was often flat, leading to unreliable
estimates of shape and much worse performance of the associated optimal for Pareto QOR
intervals.  To summarize, as a result of Study 2, in the context of
data with known support $[0,+\infty )$  we recommend using the optimal QOR for lognormal
or the Pareto($a=1$) method for quantile interval estimation. We take Method A in this context to be the optimal QOR for lognormal. We do not discuss Method B further in this
paper.

\subsection{Study 3:  Comparison of various intervals estimators based on QOR for generalized lambda models}\label{study3}

In this Section we compare the optimal QOR for the Cauchy and/or lognormal (Method A) with Methods C,D of \cite{SU09} and our Method E, adaptive QOR-GLD for $u\in [0.05,0.95]$. We consider estimation for the standard Cauchy(0,1), the heavy-tailed G-H(0.2, 0.2) \citep[see][]{TU77,HO85,hks-2008}, (also considered by \cite{SU09} for $u=0.05,0.25,0.50,0.75$ and $0.95$), the standard lognormal LN(0,1) and the exponential EXP(1) distributions for sample sizes $n=100$ and 500.  We make our comparisons with empirical coverage probabilities from 10,000 simulated runs for each choice of $u$.  Widths of these intervals are very similar and are therefore not reported.

For estimation of the GLD parameters, we use  maximum likelihood estimates for the FMKL parameterization \cite[e.g.][]{SU07}.  While many other estimators of the GLD parameters are available, after making several comparisons \cite{CO&ME15} find that the MLE estimators are a good choice since they often result in decreased estimator variability.  However, they note that there are ongoing  improvements to GLD parameter estimation so that the results that follow could later be improved with the introduction of better estimators.

\subsubsection{Single population quantiles}

\begin{figure}[h!t]
\centering
\includegraphics[scale = 0.8]{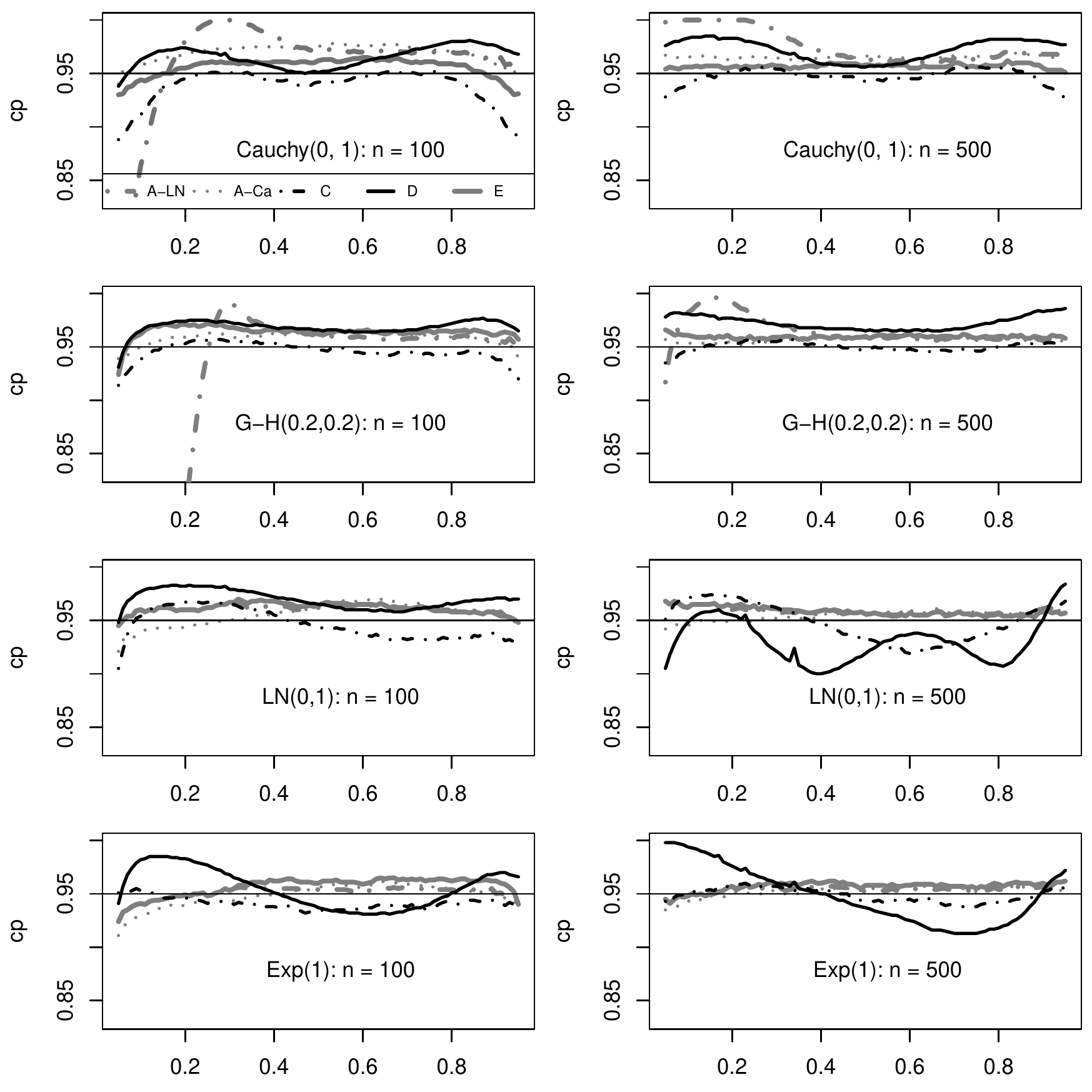}
\caption{Simulation results for coverage probability for the Cauchy, G-H$(0.2,0.2)$, LN$(0,1)$ and EXP$(1)$ distributions, plotted as a function of $u$, for samples sizes $n=100$ (left) and 500 (right).}\label{figure5}
\end{figure}

The simulated coverage probabilities for the various interval estimators of the quantiles are depicted in Figure \ref{figure5} for  $u=0.05:0.95/0.01$.  For the QOR approach to choosing the optimal bandwidth, we include the QOR for the lognormal (A-LN; thick grey dot-dash), the Cauchy (A-Ca; dot grey) and the adaptive GLD (E; thick grey line).  We also consider Methods C (C; black dot-dash) and  (D; black line) of \cite{SU09}.  For the Cauchy and lognormal distributions we expect the Cauchy QOR and lognormal QOR, respectively, to have suitable bandwidths that achieve close to nominal coverage and this is indeed the case for both sample sizes $n=100$ and $500$ for all $u$.  However, it can be seen from the coverage plots for the G-H distribution that a poorly chosen QOR can result in inadequate coverage.  The lognormal QOR when used for the symmetric Cauchy and the approximately symmetric G-H$(0.2,0.2)$ distribution results in very poor coverage for small $u$ and $n=100$ and coverage that is too conservative when $n=500$ and for $u < 0.4$.  Surprisingly, the optimal Cauchy QOR typically performs well for both the symmetric and non-symmetric distributions.

 While we expect the Cauchy QOR to work well for symmetric distributions, its simplicity and performance for these non-symmetric distributions means that it may be an attractive option in general.  However, rather than explore the Cauchy QOR further, it should be pointed out Method~E, the adaptive GLD-QOR (thick grey line), performs remarkably well for all distributions considered and for both $n=100$ and $n=500$.  Given that the GLD can approximate many distributions, these results suggest that in comparison to the other QOR approaches, it provides an excellent means to obtain a good bandwidth for the problem at hand.  It should also be noted that GLD only needs to approximate the true underlying distribution reasonably well in order to find a good bandwidth and consequently good coverage.  This is not necessarily the case other methods based on the GLD and we will comment on this next.

While Methods C and D are often slightly conservative for $n=100$, the performance of these methods may suffer for larger $n$.  As noted by \cite{SU09}, their performance  depends on the ability of the GLD to approximate the true underlying distribution.  Consequently, because bias in the density estimation does not diminish with increasing sample size, it can lead to poor coverage, especially when $n$ is large.

\begin{figure}[h!t]
\centering
\includegraphics[scale = 0.8]{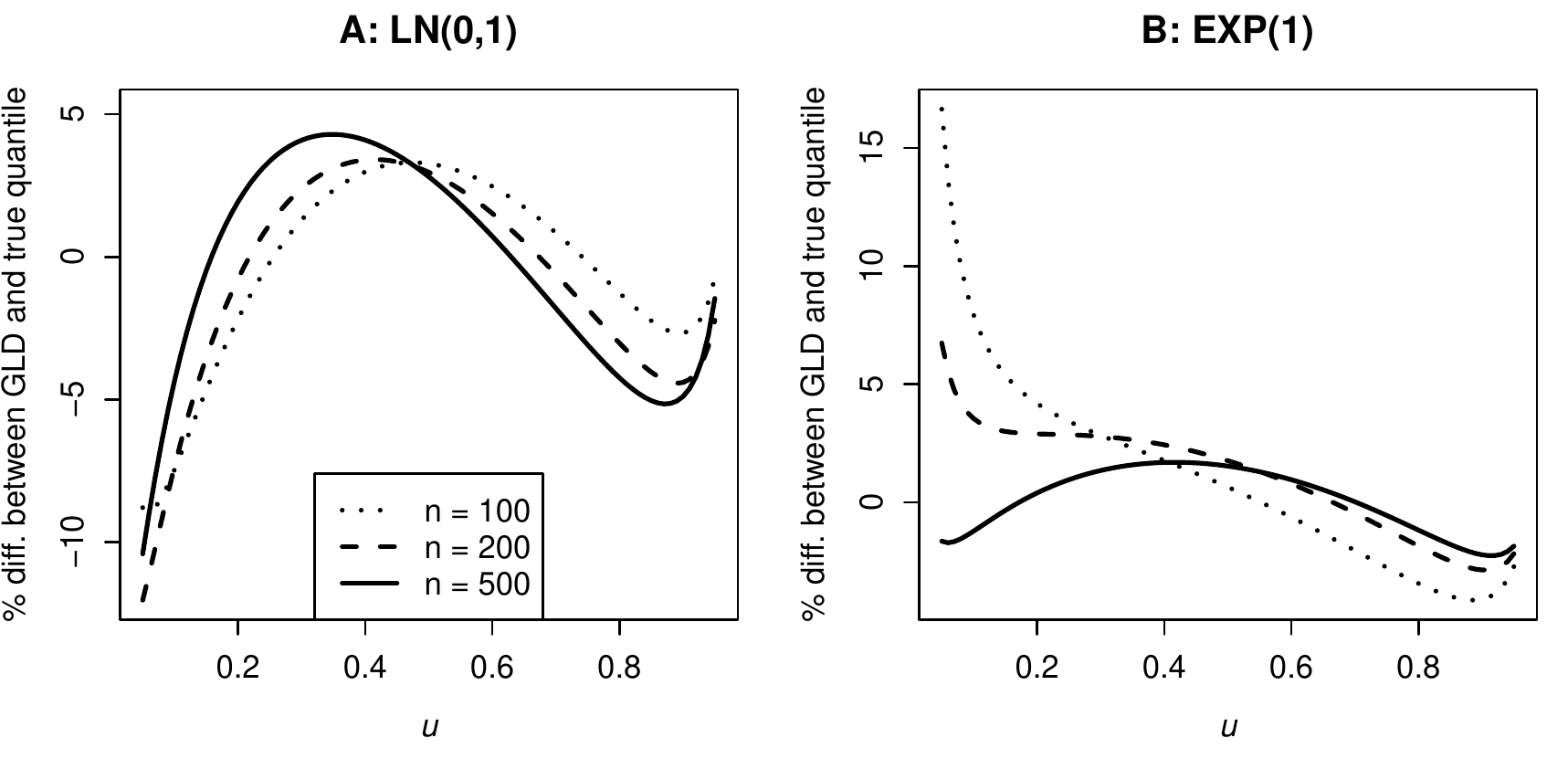}
\caption{Differences (as a percentage) between simulated GLD quantiles and the actual quantiles for the true distribution. The LN$(0,1)$ and EXP$(1)$ distributions are considered for three sample sizes, 100, 200 and 500; in this study there are 1000 replications.}\label{figure6}
\end{figure}

To assess the ability of the GLD distribution to adequately approximate other distributions, we  compare the GLD quantiles with the quantiles from the distributions that it is seeking to approximate. In Figure \ref{figure6} we plot the differences between simulated GLD quantiles versus the actual quantiles for the LN$(0,1)$ and EXP$(1)$ distributions for varying $u$ and three choices of sample sizes.  The differences are reported as a percentage difference with a negative value indicating an under-approximation for the GLD quantile.  The simulated GLD quantile was obtained over 1000 simulation runs.  For the LN$(0, 1)$ distribution, it can be seen that the bias in the GLD does not necessarily decrease as $n$ increases.  For the EXP$(1)$ distribution, while the bias does decrease for increasing $n$, it is still evident for $n=500$. Given that Methods C and D are assuming that the GLD is a consistent estimator for the true underlying density, the persistent bias for some distributions and with large $n$ can explain why these methods can return poor coverage as $n$ increases.

\begin{figure}[h!t]
\centering
\includegraphics[scale = 0.8]{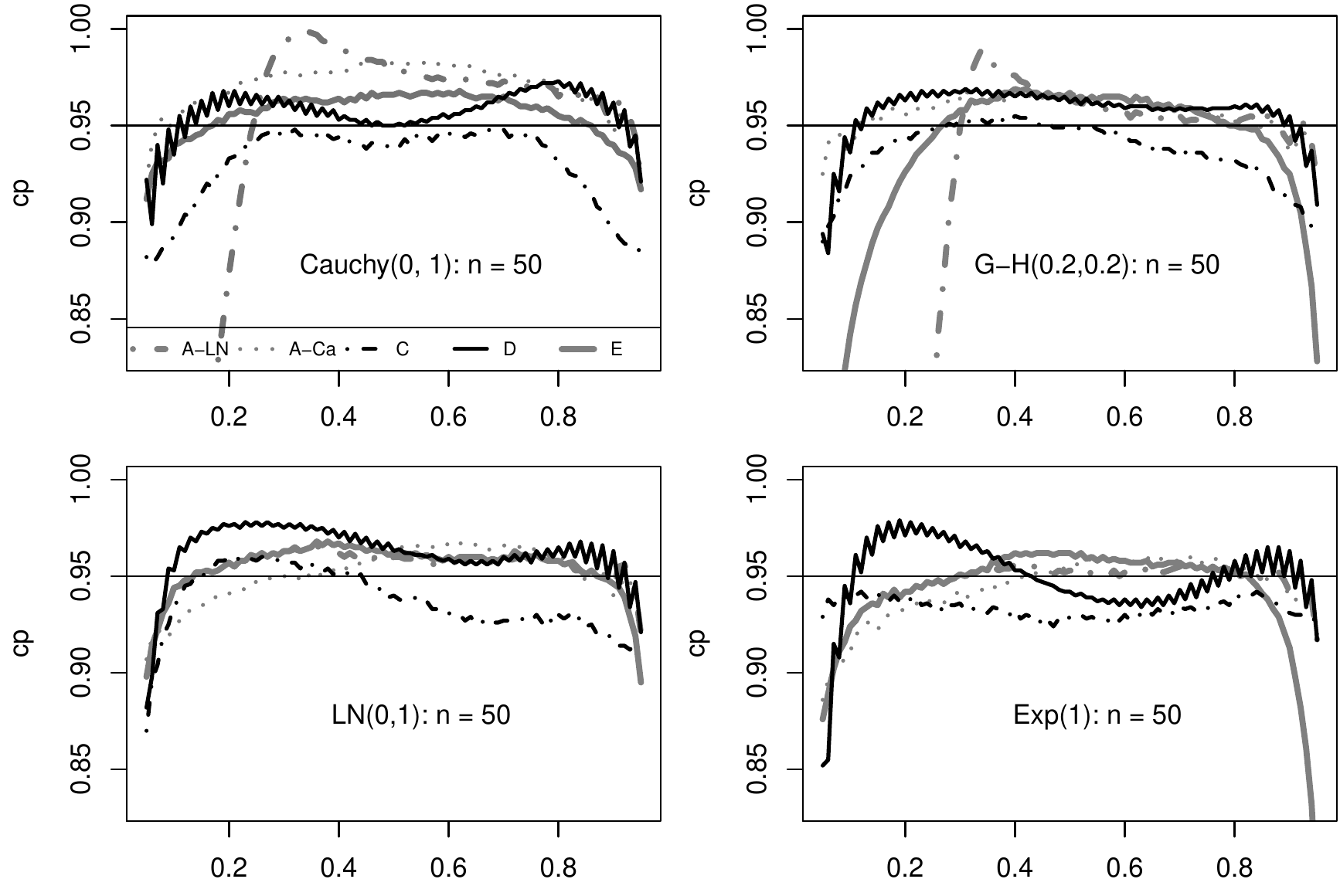}
\caption{Simulation results for coverage probability for the Cauchy$(0,1)$, G-H$(0.2,0.2)$, LN$(0,1)$ and EXP$(1)$ distributions for varying $u$ and samples size $n=50$.}\label{figure7}
\end{figure}

Nevertheless, our investigations suggest that Methods C and D are comparatively very good interval estimators for smaller $n$.  In Figure \ref{figure7} we depict the simulated coverages for several methods when the sample size is just $n=50$.  While the QOR approaches perform reasonably well provided that $u$ is neither small or large, the coverage can otherwise be very poor.  This is understandable for small $n$ where direct estimation of the quantile density is being attempted in regions with very few observations.  On the other hand, Methods C and D are comparatively quite good across a wide range of $u$.  For these methods, replacing the unknown density with the estimated GLD density works well since the quantile density is not heavily effected by sparsity of points in some regions of $u$.  That is, all 50 observations are used in the estimation of the four GLD parameters and provided that these estimates are reasonably accurate, then the methods should perform well with small $n$.

\subsubsection{Extension to comparing quantiles from two-populations}\label{section:diffs}

 Generalizing \eqref{ciquant} it is straightforward to construct large sample interval estimators for linear combinations of quantiles from independent populations.  For example, let $y_p$ denote the $p$th quantile for a new population where $0 < p < 1$ and denotes its estimator $\hat{y}_p$ based on a random sample of size $m$.  Since $\hat{x}_u$ and $\hat{y}_p$ are independent, then a large sample confidence interval for $x_u-y_p$ is
 \begin{equation}\label{int_diff}
 (\hat{x}_u-\hat{y}_p)\pm z_{1-\alpha/2}\sqrt{\frac{\tau_u^2}{n}+\frac{\nu^2_p}{m}}
 \end{equation}
 where $m\text{Var}[\hat{y}_p]\doteq \nu ^2_p$.

\begin{figure}[h!t]
\centering
\includegraphics[scale = 0.8]{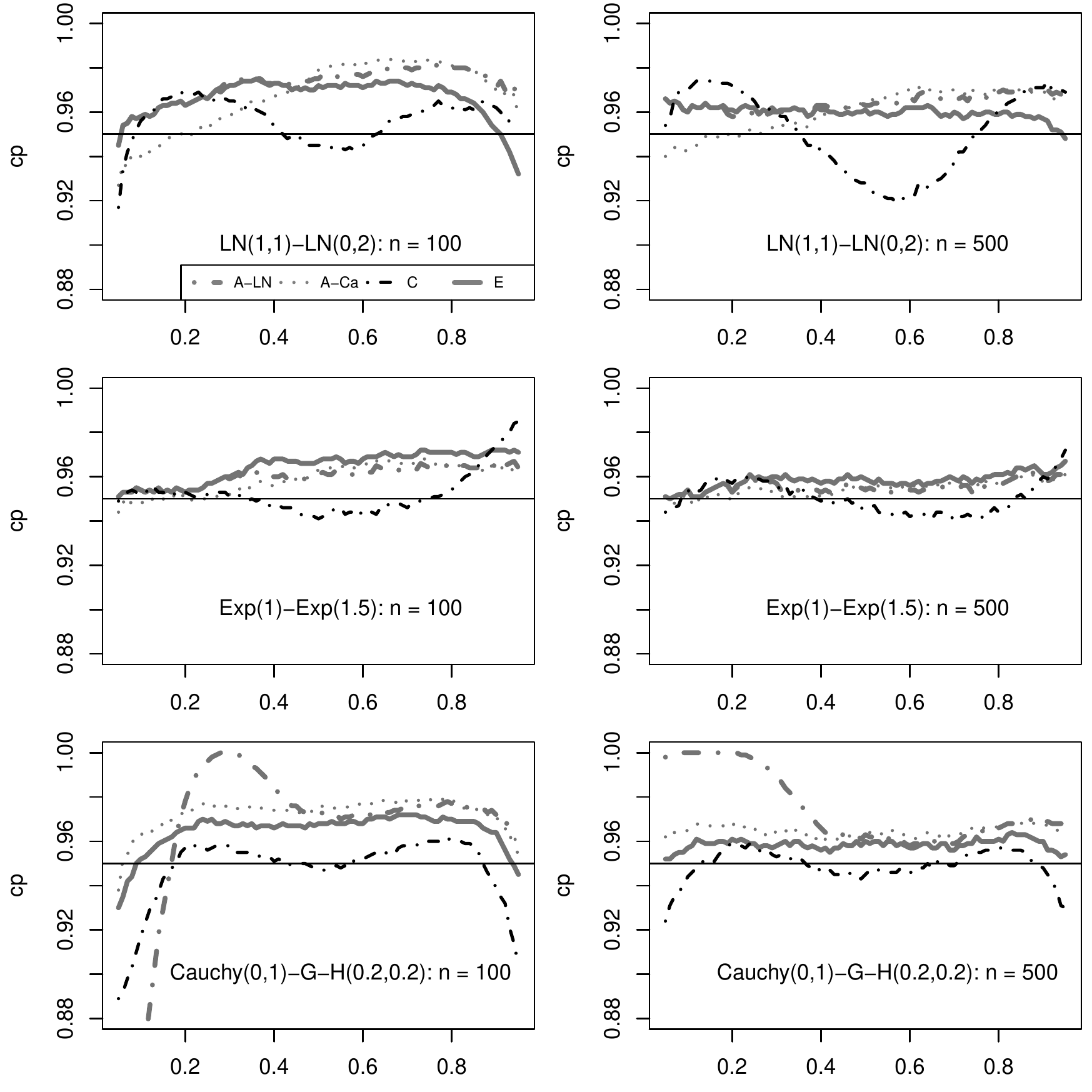}
\caption{Simulation results for coverage probability for the differences in quantiles from several distributions distributions for varying $u$ and samples sizes $n=100$ (left) and $n=500$ (right).}\label{figure8}
\end{figure}

In Figure \ref{figure8} we provide simulated coverage probabilities for for Methods A (with lognormal and Cauchy QOR) and the GLD Methods C and E.  With the exception of when using the lognormal QOR for symmetric (or close symmetric distributions as is the case for the G-H(0.2, 0.2) distribution) the methods typically return good coverage for most choices of $u$.  In particular, Method E which used the GLD QOR provides the most stable coverage with coverage probabilities rarely below nominal and usually between 0.95 and 0.97 for both choices of $n$.

\subsection{Study 4: Methods F,G and H}\label{study4}

Methods F,G and H are due to \cite{SO12}
and their simulation studies for sample sizes 200 from the exponential (their Table~3.1) and GLD$(1,-1,-1/8,-1/8)$
(their Table~3.2), comparing the MSE of the following estimates F,G, and H of $q(u)$ for both Epanechnikov
and triangular kernels, with 3 choices of constant in $u$ bandwidth $h_n=0.15, 0.19$ and $0.25.$  Neither kernel gave uniformly better results and the bandwidth results are not that dependent on choice of constant. Their Table 3.1 shows that their Method H has smaller MSE than Methods F and G for $u=0.2, 0.4, 0.6, 0.8$ and 0.9. We tried to replicate
their results but obtained only partial agreement that their $H$ is preferable to $F$ in terms of MSE. We also included
Method E, the adaptive QOR for the  GLD distribution for comparison, and found it competitive for this sample size
in the context of exponential data.   More importantly, for larger $n=500$ say, the bias and MSE of their Method F estimator is larger than for $n=200$ when $u=0.8$ or $u=0.9$. This suggests that with constant bandwidth the bias of all their
estimators will grow without bound, leading to small coverage probabilities of the associated confidence intervals.
 One would expect that for all of their methods F,G and H a bandwidth of order $n^{-1/5}$ is required for minimizing
the MSE, but there is no discussion of bandwidth dependence on $n$ in their paper, although it is denoted $h_n$. 

\subsection{Study 5:  Example of Heart-rate data}\label{study5}

\cite{SU09} compared heart rate observations for males with those for females where there are 65 observations in each group.  The data set does not explicitly state which data belongs to which group so we will follow Su's lead and refer to simply gender 1 and gender 2.  Since the maximum likelihood estimates for the RS GLD parameterization resulted in the best QQ plots, it was this parameterization that was used and that we will subsequently use here.  Su reported that the modes for the fitted distributions are located at the 0.48 and 0.64 quantiles for gender 1 and gender 2.  Based on the parameter estimates reported by \cite{SU09}, we were able to replicate the reported 95\% confidence intervals for the population quantiles given as $[71.02,\ 74.81]$ and $[74.89,\ 80.33]$ respectively.  As noted by Su, the lack of overlap for these intervals indicates a significant difference between the quantiles.

A potential problem with the reliance on separate confidence intervals to determine a difference between two independent groups is in what to do conclude when they do overlap.  Overlapping confidence intervals can still indicate a significant difference and the intervals reported above were, in fact, very close to overlapping.  We will instead use the QOR approach based on the GLD to find an interval for the difference in quantiles (see Section \ref{section:diffs}) as well as separate intervals for each group.  The QOR for the RS GLD can be found in Table \ref{table:QOR}.  For gender 1 and gender 2 our separate 95\% intervals are $[70.94,\ 75.06]$ and $[75.30,\ 81.00]$ respectively which are similar to those reported above.  For the difference between the quantiles between gender 2 and gender 1, our 95\% confidence interval is $[1.63,\ 8.67]$ which suggests a positive difference.

\begin{table}[ht]
\centering
\begin{tabular}{cccccc}
  \hline
   \multicolumn{2}{c}{Analytical}  && \multicolumn{3}{c}{QOR} \\
  gender 1 & gender 2 && gender 1 & gender 2 & gender 2 $-$ gender 1 \\ \cmidrule{1-2} \cmidrule{4-6}
   (70.44, 75.42) & (73.95, 81.08) & & (70.29, 75.71) & (74.4, 81.89) &  (0.52, 9.77)\\
  \hline
\end{tabular}
\caption{99\% confidence intervals for the quantiles for gender 1 and gender 2 (for the the analytical and QOR methods) and for the difference between the gender 2 and gender 1 quantiles using the QOR method.}\label{table:99gender}
\end{table}

To emphasize our point about the advantages of being availed a single interval for the difference in quantiles, we increase the confidence level to 99\% and report the intervals in Table \ref{table:99gender}.  The intervals are, as expected, wider and the gender 1 and gender 2 intervals now overlap for both methods.  However, the 99\% confidence interval for the difference in quantiles using the QOR approach is free of zero and still indicative of a difference between the groups.

\section{Summary and discussion}\label{summary}

A kernel estimator of the quantile density $q(u)$ that has been the subject of investigation by several authors is known to have optimal, (in the sense of minimizing the asymptotic
mean squared error at $u$), bandwidth that depends on the underlying location-scale family only through the quantile optimality ratio  QOR$(u)=q(u)/q''(u).$  By examining numerous examples of this ratio, we found it to be relatively well behaved (compared to $q$ and its derivatives) with graph similar in shape to the square of the density quantile $f(x_u)$. The consequence for estimation of $q(u)$ needed in construction of the simple confidence intervals (\ref{ciquant}) is that $q''(u)$  need not be estimated. Rather, a representative QOR that is optimal for one family turns out to be more than adequate for many similarly shaped families. We called the  representative bandwidth approach Method~A.

In particular for symmetric unimodal distributions with infinite support,
we found that using the optimal for Cauchy QOR bandwidth led to relatively good coverage
for all $u\in [0.05,0.95]$ and $n\geq 400$. For smaller $n$, the good coverage range of $u$
diminishes, but is still competitive with most intervals.
For unimodal skewed-right distributions with support on $[0,\infty)$ a similar result holds: \  the
optimal for lognormal QOR (or Pareto($a=1$) model) provides a bandwidth leading to very good coverage for $u\in [0.05,0.95]$ for a variety of commonly assumed distributions. In this case a
very simple boundary correction is required for the bandwidth.  Method~A quantile density estimates may also improve the performance of ratios of linear combinations of quantiles such as  the standardized median \cite{S-2013b} or skewness measures \cite{S-2014}.
In those papers a truncated kernel density estimator for $f(x_u)$ was the basis for
estimating its reciprocal $q(u)$, but a small simulation study shows that the direct estimators of $q(u)$
proposed herein are superior, especially for small or large $u$.

 Method~B was less
successful. It tried assuming much less, such as a symmetric Tukey $\lambda $ model,
first estimating the parameter and then using optimal QOR$_{\hat \lambda }(u).$  We also
tried one and two-parameter Pareto models, but finding the optimal QOR for the estimated
model appears to be a more complicated procedure with less success.

Some researchers are willing to assume the four-parameter generalized lambda family of
distributions. We found that by estimating the parameters first and then using  the optimal QOR for this
GLD led to relatively good coverage, compared to Methods C and D of \cite{SU09},
 of confidence intervals for $n\geq 100.$ We called this
Method E. We showed that Methods C and D are unlikely to be useful for larger $n$ if
the underlying distribution was not in the GLD, because the quantiles of the estimated
GLD do not converge to the true quantiles. Nevertheless, for small $n$ we recommend Method~C, which has remarkably good coverage over a large range of $u$. Further research into parameter estimation for the GLD could lead to improvements in Method E.

The distribution-free Methods F,G and H \cite{SO12} which use a constant bandwidth were found uncompetitive.  The only constant bandwidth QOR that we could find arose from a
bimodal distribution with finite support, see the Appendix in Section \ref{appendix}.

Finally, we showed that it is easy to extend these results to two-sample problems
where it is desired to compare quantiles from independent samples.  For the heart-rate
data of \cite{SU09}, we found a single interval for the difference in quantiles which is
easier to carry out tests for a significant difference in quantiles, rather than relying
on the more complicated test based on separate confidence intervals.

In summary, examination of the QOR for many parametric families shows that the optimal
bandwidth based on it need not be very accurate to obtain a good estimate of the quantile
density required for simple distribution-free confidence intervals. At the same time the
shape of the QOR cannot be ignored if one desires good coverage of confidence intervals
for quantiles, for moderate and large sample sizes.

With regard to further research, an  important class of models for which this methodology may prove useful is the  mixtures of normal distributions, for which one can easily obtain parameter estimates using existing software packages. One would need an expression for, and an estimate of, the QOR in this case, in order to obtain interval estimates for quantiles.


\section{Appendix.  Additional examples of the QOR}\label{appendix}

The distributions below are described in \cite{J-K-B-1994} and \cite{J-K-B-1995}.

\subsubsection*{Cauchy.} For $F(x)=0.5+\arctan (x)/\pi$ the quantile function is $x_u=Q(u)=\tan (\pi (u-0.5)).$
Hence
\begin{eqnarray} \nonumber
  q(u) &=&  \pi \sec ^2\{\pi(u-0.5)\}\\ \label{cauchyqden} \nonumber
  q'(u) &=& 2\pi^2 \left [\tan \{\pi(u-0.5)\}+ \tan ^3\{\pi(u-0.5)\}\right ]\\ \nonumber
  q''(u)&=& 2\pi^3 \sec ^2\{\pi(u-0.5)\}\left [1+3\tan ^2\{\pi(u-0.5)\}\right ]\
\end{eqnarray}
  Elementary calculations show that this ratio  equals to $1/(2\pi \sqrt {3})$ times the Cauchy density having median 0 and scale parameter $1/\sqrt {3}$, evaluated at the quantile function. A formula for the QOR is given in Table~\ref{table:QOR}.

\subsubsection*{Laplace.} If $f(x)=e^{-|x|}/2$,  the quantile function is
$Q(u)=\ln (2u)$ for $0<u<1/2$ and $Q(u)=-\ln\{2(1-u)\}$ for  $1/2<u<1$, so the quantile optimality ratio is
$QOR(u)=u^2/2$ for $0<u<1/2$ and $(1-u)^2/2$ for $1/2<u<1;$ its graph is shown in the upper right plot of Figure~\ref{fig1}. Elementary calculations show that  $QOR(u)=f_{1/2}(x_u)/8$, a scale multiple of the
Laplace distribution with median 0 and scale 1/2, evaluated at the quantile function.

\subsubsection*{Logistic.} The density function $f(x)=e^{-x}\{1+e^{-x}\}^2$ for all $x$ and the quantile
function $Q(u)=\ln(u)-\ln(1-u), $ for $0<u<1$.  It follows after taking its derivatives that the QOR is given by $2\{(1-u)^3-u^3\}/\{ u(1-u)\}^2.$

\subsubsection*{Bimodal with constant ratio.} Since at least one publication \cite{soni-2012} used a constant ratio
in the definition of the optimal bandwidth (\ref{bwconstant}), we looked for a density $f$ which
leads to a constant QOR. One possibility is
   $f(x)=\{2(e-|x|)\}^{-1}$ for all $|x|< e-1.$ It has constant QOR$(u)=1/4.$   However when the QOR in the  bandwidth
is constant the ensuing estimator of $q(u)$ is poor for most $u$
compared to many others, so we did  not consider it further.

\subsubsection*{Tukey $\lambda $.}  Recall that the Tukey$(\lambda )$ distributions have quantile density function given by $q(u)=u^{\lambda -1}+(1-u)^{\lambda -1}$ for all $\lambda .$ There is no closed form for the density $f$ itself except when $\lambda =0$ and the
distribution $F$ is logistic, or when $\lambda =1$ or 2 and the distribution is uniform.
When $\lambda =2.5,$ the quantile optimality ratio $q(u)/q''(u)$ is approximately constant over the range $0.1\leq u \leq 0.9$.

\subsubsection*{Pareto.}

For $x>0$ let $F(x;a)=1-(1+x)^{-a}$, where $a>0 $ is a shape parameter; such distributions are called Lomax or Pareto Type II. The quantile function $x_u=Q(u)=(1-u)^{-1/a}-1$, for $0<u<1$; and, the quantile optimality ratio
QOR$(u)=a^2(1-u)^2/\{(1+a)(1+2a)\}.$

\subsubsection*{Gamma.}
 Recall the gamma distribution has density $f(x;\alpha)=x^{\alpha -1}e^{-x}/\Gamma (\alpha )$ for $x>0$, where $\alpha >0$ is the shape parameter and  $\Gamma(\alpha )$ is the gamma function. Using (\ref{eqn:quantilederivs}) and the fact that $f'(x;\alpha)=g(x)f(x;\alpha)$, where $g(x)=(\alpha -1)/x-1$, one finds the quantile optimality ratio:
\begin{equation} \label{gammaqor}
\text{QOR}_\alpha (u) =\frac {x_u^2\,f^2(x_u;\alpha )}{(2\alpha -1)(\alpha-1)-4(\alpha-1)x_u +2x_u^2}  ~.
\end{equation}
Note that this ratio is negative for some $u$ if $0.5<\alpha <1.$.

\subsubsection*{Weibull.}

For $\beta >0$ the Weibull distribution function is defined for each $x$ by $F(x;\beta)=1-
e^{-x^\beta }$. Clearly $f'(x;\beta)=g(x)f(x;\beta)$, where $g(x)=(\beta -1)/x-\beta /x^{\beta -1}.$ Again applying (\ref{eqn:quantilederivs}) yields the result:
\begin{equation} \label{weibullqor}
\text{QOR}_\beta (u) =\frac {x_u^2\,f^2(x_u;\beta )}
{(2\beta -1)(\beta-1)-3\beta(\beta-1)x_u^\beta  +2\beta ^2 x_u^{2\beta }}  ~.
\end{equation}
 The special case  of the Weibull when $\beta =1$ is of interest because its ratio  $QOR(u)=(1-u)^2/2, $ which is the same form as the ratio for the Pareto II family.  However, for any $a>0$ the coefficient of $(1-u)^2$ for the Pareto~II $(a)$ distribution
is less than 1/2.

\end{document}